\documentclass[conference,a4paper]{IEEEtran}

\newif\ifauthorversion
\authorversiontrue

\newif\ifreview
\reviewfalse

\newif\iffinal
\finaltrue

\usepackage{amsmath,amssymb,amsfonts}
\usepackage{algorithmic}
\usepackage{graphicx}
\usepackage{textcomp}
\usepackage{xcolor}
\def\BibTeX{{\rm B\kern-.05em{\sc i\kern-.025em b}\kern-.08em
    T\kern-.1667em\lower.7ex\hbox{E}\kern-.125emX}}

\usepackage{algorithmic}
\usepackage{textcomp}
\usepackage{tabularx}
\newcolumntype{Y}{>{\centering\arraybackslash}X} %
\usepackage{xcolor}
\usepackage[inline]{enumitem}
\usepackage{dsfont}
\usepackage{xspace}
\usepackage{multirow}
\usepackage{pifont}

\usepackage{hyperref} %
\usepackage{tikz}
\usepackage{orcidlink}
\usepackage[binary-units]{siunitx} %
\newlist{questions}{enumerate}{1}
\setlist[questions,1]{label=\textbf{RQ\arabic*.},ref=\textbf{RQ\arabic*}}
\usepackage[section,numberedsection=autolabel,nogroupskip,nonumberlist]{glossaries}
\newif\ifprintglossary
\printglossarytrue

\ifprintglossary
    \makeglossaries  %
    
\else
    \glsdisablehyper
\fi

\newacronym{aae}{AAE}{Adversarial Autoencoder}
\newacronym{ae}{AE}{Autoencoder}
\newacronym{bce}{BCE}{Binary Cross Entropy}
\newacronym{ce}{CE}{Cross Entropy}
\newacronym{cnn}{CNN}{Convolutional Neural Network}
\newacronym{dcgan}{DCGAN}{Deep Convolutional Generative Adversarial Network}
\newacronym{dl}{DL}{Deep Learning}
\newacronym{dp-stg}{DP-STG}{Differentially Private Synthetic Trajectory Generator}
\newacronym{dp}{DP}{Differential Privacy}
\newacronym{dtw}{DTW}{Dynamic Time Warping}
\newacronym{em}{EM}{Exponential Mechanism}
\newacronym{emd}{EMD}{Earth Mover's Distance}
\newacronym{exGAN}{exGAN}{Except-Condition GAN}
\newacronym{fs}{FS-NYC}{Foursquare NYC~\cite{fs_nyc}}
\newacronym{gan}{GAN}{Generative Adversarial Network}
\newacronym{geo-ind}{Geo-Ind}{Geo-Indistinguishability}
\newacronym{gru}{GRU}{Gated Recurrent Unit}
\newacronym{hd}{HD}{Hausdorff Distance}
\newacronym{jsd}{JSD}{Jensen Shannon Distance}
\newacronym{lldp}{LLDP}{Local Label Differential Privacy}
\newacronym{ldp}{LDP}{Local Differential Privacy}
\newacronym{lstm}{LSTM}{Long Short-Term Memory}
\newacronym{mae}{MAE}{Mean Absolute Error}
\newacronym{mi}{MI}{Mutual Information}
\newacronym{mia}{MIA}{Membership Interference Attack}
\newacronym{swd}{SWD}{Sliced Wasserstein Distance}
\newacronym{ttd}{TTD}{Total Travelled Distance}
\newacronym{mlp}{MLP}{Multi-Layer Perceptron}
\newacronym{mse}{MSE}{Mean Squared Error}
\newacronym{mwe}{MWE}{Minimal Working Example}
\newacronym{ntg}{NTG}{Noise-TrajGAN}
\newacronym{rtct}{RTCT}{Reversible Trajectory-to-CNN Transformations}
\newacronym{trr}{TRR}{Time Reversal Ratio}
\newacronym{ttur}{TTUR}{Two Time Update Rule}
\newacronym{nlp}{NLP}{Natural Language Processing}
\newacronym{poi}{POI}{Point of Interest}
\newacronym{poc}{PoC}{Proof-of-Concept}
\newacronym{RAoPT}{RAoPT}{Reconstruction Attack on Protected Trajectories}
\newacronym{rdp}{RDP}{Rényi Differential Privacy}
\newacronym{relu}{ReLU}{Rectified Linear Activation}
\newacronym{rnn}{RNN}{Recurrent Neural Network}
\newacronym{sdd}{SDD}{Sampling Distance and Direction}
\newacronym{sota}{SOTA}{State Of The Art}
\newacronym{stc}{STC}{Spatial-Temporal-Categorical Distance}
\newacronym{stg}{STG}{Synthetic Trajector Generator}
\newacronym{stn}{STN}{Large Spatial Transformer Network}
\newacronym{tanh}{tanh}{Hyperbolic Tangent}
\newacronym{TRA}{TRA}{Trajectory Reconstruction Attack}
\newacronym{TSG}{TSG}{Two-Stage-\gls{gan}}
\newacronym{tul}{TUL}{Trajectory User Linking}
\newacronym{uop}{UoP}{Unit of Privacy}
\newacronym{vae}{VAE}{Variational Autoencoder}
\newacronym{wd}{WD}{Wasserstein Distance}
\newacronym{tcac}{TCAC}{Trajectory Category Auxiliary Classifier}
\newglossaryentry{conv1d}{
    name={Conv1D},
    short={Conv1D},
    description={1D Convolution},
    first={Conv1D},
    long={Conv1D}
}
\newglossaryentry{fc}{
    name={FC},
    short={FC},
    description={Fully Connected layer. Also called \textit{Dense} or \textit{Linear} layer},
    first={Fully Connected (FC)},
    long={Fully Connected}
}
\newglossaryentry{dp-sgd}{
    name={DP-SGD},
    short={DP-SGD},
    first={Differentially Private Stochastic Gradient Descent (DP-SGD)},
    long={Differentially Private Stochastic Gradient Descent},
    description={Differentially Private Stochastic Gradient Descent, refer Section~\ref{sec_dp-sgd_bg}}
}
\newglossaryentry{wgan}{
    name={WGAN},
    first={WGAN},
    description={Wasserstein \gls{gan}\cite{wgan}}
}
\newglossaryentry{wgan-lp}{
    name={WGAN-LP},
    first={WGAN-LP},
    description={\gls{wgan} with Lipschitz Penalty \cite{wgan_lp}}
}
\newglossaryentry{wgan-gp}{
    name={WGAN-GP},
    first={\gls{wgan} with Gradient Penalty (WGAN-GP)},
    description={\gls{wgan} with Gradient Penalty \cite{iWGAN}}
}
\newglossaryentry{mnist-seq}{
    name={MNIST-Seq},
    first={MNIST Sequential (MNIST-Seq)},
    description={MNIST Sequential Dataset: Images are transformed to sequences of length $28$ with $28$ features each \cite{Esteban2017}}
}
\newglossaryentry{ltg}{
    name={LSTM-TrajGAN},
    short={LSTM-TrajGAN}
    first={LSTM-TrajGAN},
    description={LSTM-TrajGAN~\cite{Rao2020}}
}
\newcommand{\lstmtrajgan}{\gls{ltg}\xspace}

\makeatletter\let\myfsize\f@size\makeatother
\newcommand{\subheading}[1]{
    \noindent{\textbf{#1.}}
    \addcontentsline{toc}{subsubsection}{#1}
}
\iffinal

\else

\fi

\newcommand{\probP}{\text{I\kern-0.15em P}}  %
\newcounter{goal}
\renewcommand*\thegoal{G\arabic{goal}}
\newcommand{\gref}[1]{\hyperref[#1]{\textbf{\ref*{#1}: \csname goalname#1\endcsname}}\xspace}

\newcommand{\goal}[2]{%
    \refstepcounter{goal}%
    \textbf{\thegoal: #1}\label{#2}%
    \addcontentsline{toc}{subsection}{\thegoal: #1}%
    \expandafter\gdef\csname goalname#2\endcsname{#1}%
}

\newcommand{\refer}{ref.\ }

\newtheorem{definition}{Definition}

\begin{document}

\title{Synthetic Trajectory Generation Through Convolutional Neural Networks
}
\author{
    \IEEEauthorblockN{
        Jesse Merhi\IEEEauthorrefmark{1}\IEEEauthorrefmark{3}\orcidlink{0009-0009-0562-2128},
        Erik Buchholz\IEEEauthorrefmark{1}\IEEEauthorrefmark{2}\IEEEauthorrefmark{3}\orcidlink{0000-0001-9962-5665},
        Salil S. Kanhere\IEEEauthorrefmark{1}\orcidlink{0000-0002-1835-3475}
    }
    \IEEEauthorblockA{
        \IEEEauthorrefmark{1} \textit{University of New South Wales},
        Sydney, Australia
    }
    \IEEEauthorblockA{
        \IEEEauthorrefmark{2} \textit{CSIRO's Data61, Cyber Security CRC}
    }
    \IEEEauthorblockA{
        \IEEEauthorrefmark{3} \textit{Both authors contributed equally to this research.}
    }
}

\ifauthorversion
    \newcommand\copyrighttext{%
      \footnotesize \textcopyright 2024 IEEE. To appear in the proceedings of the 21st Annual International Conference on Privacy, Security \& Trust (PST 2024).
      This is the author’s version of the work. It is posted here for your personal use. Not
for redistribution. 
      }
    
    \newcommand\copyrightnotice{%
    \begin{tikzpicture}[remember picture,overlay]
    \node[anchor=south,yshift=50pt] at (current page.south) {\fbox{\parbox{\dimexpr0.75\textwidth-\fboxsep-\fboxrule\relax}{\copyrighttext}}};
    \end{tikzpicture}%
    }
\fi

\maketitle

\begin{abstract}
Location trajectories provide valuable insights for applications from urban planning to pandemic control.
However, mobility data can also reveal sensitive information about individuals, such as political opinions, religious beliefs, or sexual orientations.
Existing privacy-preserving approaches for publishing this data face a significant utility-privacy trade-off.
Releasing synthetic trajectory data generated through deep learning offers a promising solution.
Due to the trajectories' sequential nature, most existing models are based on recurrent neural networks (RNNs).
However, research in generative adversarial networks (GANs) largely employs convolutional neural networks (CNNs) for image generation.
This discrepancy raises the question of whether advances in computer vision can be applied to trajectory generation.
In this work, we introduce a Reversible Trajectory-to-CNN Transformation (RTCT) that adapts trajectories into a format suitable for CNN-based models.
We integrated this transformation with the well-known DCGAN in a proof-of-concept (PoC) and evaluated its performance against an RNN-based trajectory GAN using four metrics across two datasets.
The PoC was superior in capturing spatial distributions compared to the RNN model but had difficulty replicating sequential and temporal properties.
Although the PoC's utility is not sufficient for practical applications, the results demonstrate the transformation's potential to facilitate the use of CNNs for trajectory generation, opening up avenues for future research.
To support continued research, all source code has been made available under an open-source license.
\end{abstract}


\ifauthorversion
    \copyrightnotice
\fi

\section{Introduction}\label{sec_intro}

Due to the omnipresence of sensor-equipped devices like smartphones, large quantities of location data are collected daily.
These trajectory datasets offer the potential for various use cases, such as public transport optimisation and pandemic control.
However, location data yields severe privacy implications as it allows the re-identification of individuals~\cite{DeMontjoye2013, Rossi2015} and the inference of personal attributes~\cite{Primault2019, Abul2008}.
To highlight the high information content of location traces, De Montjoye et al.~\cite{DeMontjoye2013} managed to identify \SI{95}{\%} of users in a mobile phone dataset based on four locations only.
Moreover, researchers determined Islamic taxi drivers in an anonymised dataset by correlating the mandatory prayer times with their breaks \cite{Franceschi-Bicchierai}.

Researchers have proposed various methods for the privacy-preserving release of trajectory datasets~\cite{Primault2019, Errounda2020, Jin2021, Miranda-Pascual2023}.
These methods focus on syntactic privacy, such as $k$-anonymity~\cite{Sweeney2002}, and semantic privacy, notably \gls{dp}~\cite{Dwork2013}.
\gls{dp} has become the de facto standard due to its strong formal guarantees.
However, these methods involve a privacy-utility trade-off, requiring data perturbation to ensure privacy~\cite{Primault2014}.
Current approaches often fail to maintain sufficient utility in the protected data for effective analysis~\cite{Rao2020, Qu2020, Ma2021}.
Inappropriate protection can also lead to structural anomalies, such as cars not adhering to roads or ships traversing land~\cite{Naghizade2020, RAoPT, GeoPointGAN, Miranda-Pascual2023}.
These anomalies may facilitate reconstruction attacks, thus reducing the achieved privacy level~\cite{Shao2020, RAoPT}.

For these reasons, Liu et al.~\cite{Liu2018a} proposed using \glspl{gan} to generate synthetic trajectory datasets.
The \gls{dl} model retains the original dataset's key characteristics but produces inherently private, synthetic data.
Models such as \lstmtrajgan~\cite{Rao2020} show the potential utility of this method.
Yet, these solutions lack strong privacy guarantees.
There are risks of the model memorising and replicating real trajectories with minimal alterations.
To guarantee that the model does not remember training data, the usage of \glsentryshort{dp-sgd}~\cite{dpfyML} has been proposed~\cite{pets_sok}.
However, \lstmtrajgan uses a real trajectory as input, unlike standard \gls{gan} models like \gls{dcgan}~\cite{DCGAN} that use Gaussian noise.
\glsentryshort{dp-sgd} only secures training data, thus it is ineffective for architectures that rely on real data during generation~\cite{pets_sok}.

To incorporate \glsentryshort{dp-sgd} in trajectory \glspl{gan}, a \textit{noise-only} model is required.
As trajectories represent sequences of locations, \gls{rnn}-based models are the obvious choice.
However, most \gls{gan} research has centred on computer vision, focusing on \gls{cnn}-based models.
\gls{rnn}-based \glspl{gan} often train less stably than \gls{cnn} models and struggle with convergence~\cite{pets_sok}.
This raises the question of whether a \gls{cnn}-based architecture could be utilised for trajectory generation.

In other sequential domains, the move from \gls{rnn}-based models to \gls{cnn}-based models showed success.
WaveGAN~\cite{Donahue2019} adapts \gls{dcgan} for audio data generation, employing \glsentryshort{conv1d} layers and outperforming an \gls{rnn}-based model.
Similarly, PAC-GAN~\cite{pacgan} utilises a \gls{cnn}-\gls{gan} for the generation of network traffic packets where most related work had used \glspl{rnn}.
These examples highlight the potential of \gls{cnn}-based architecture's in sequential domains.

\glsreset{rtct}
This work introduces a \textit{\gls{rtct}} to facilitate the use of \gls{cnn}-based architectures for trajectory datasets.
We demonstrate its utility by integrating it with \gls{dcgan}~\cite{DCGAN} in a \gls{poc} implementation.
This study primarily focuses on the feasibility of \gls{rtct} with a \gls{cnn}-based model; we do not optimise \gls{dcgan} for trajectory generation but leave this task for future work.
We evaluate on two real-world datasets, \gls{fs}~\cite{fs_nyc} and Geolife~\cite{Geolife1} with four metrics:
\begin{enumerate*}
    \item The widely used \gls{hd}, and
    \item the sliced \gls{wd} assess the spatial distribution,
    \item \gls{ttd} the sequential properties, and
    \item a novel metric, \gls{trr}, for assesses time properties.
\end{enumerate*}
Moreover, we qualitatively analyse the generated trajectories by plotting them.

Our \gls{poc} is benchmarked against \gls{ntg}~\cite{pets_sok}, a \textit{noise-only input} variant of the \gls{sota} model LSTM-TrajGAN~\cite{Rao2020}.
\gls{ntg} was selected despite its inferior performance due to LSTM-TrajGAN's incompatibility with \glsentryshort{dp-sgd}.
After establishing a baseline comparison between our \gls{poc} and \gls{ntg}, we proceeded to train both models with \glsentryshort{dp-sgd}.

The evaluation demonstrates that our \gls{poc} more effectively captures the distribution of points within a trajectory dataset compared to the \gls{rnn} model. 
However, in terms of spatial and temporal properties measured through \gls{ttd} and \gls{trr}, respectively, \gls{ntg} outperforms our \gls{poc}.
Despite these results, the overall quality of the trajectories generated by both models does not yet meet the requirements for downstream applications.
The \gls{dp} version of our \gls{poc} achieves results comparable to \gls{ntg}.
However, qualitative analysis suggests that the noise added by \glsentryshort{dp-sgd} significantly impacts the model.
The \gls{dp} version of \gls{ntg} yields surprisingly good quantitative results.
Yet, visual inspection does not confirm effective learning.
While the results on a toy dataset were promising and showed the feasibility of integration with \glsentryshort{dp-sgd}, significant future work is required to establish a useful \gls{dp} model.

The evaluation results show that the \gls{poc} cannot reach the utility provided by (non-private) \gls{sota} models.
Yet, they highlight that the proposed transformation enables the use of \gls{cnn}-based models in trajectory generation.
Future research could explore tailored \gls{gan} designs to enhance trajectory generation.
The ultimate goal is the development of a stable \gls{cnn}-based \gls{gan} that integrates with \glsentryshort{dp-sgd} for formal privacy guarantees.
Concretely, this paper advances the field of trajectory privacy in the following ways (See Section~\ref{sec_problem_statement}):
\begin{enumerate}\glsreset{rtct}
    \item We introduce a \gls{rtct} enabling the usage of \gls{cnn} models for trajectory generation. (see Section~\ref{sec_rtct})
    \item We integrate this transformation with the well-known \gls{dcgan}~\cite{DCGAN} architecture in a \gls{poc} (see Section~\ref{sec_integration}).
    \item We evaluate our \gls{poc} on two real-world datasets, \gls{fs}~\cite{fs_nyc} and Geolife~\cite{Geolife1}, and compare it to the \gls{rnn}-based \gls{ntg} model. (see Section~\ref{sec_eval}).
    \item We integrate both models with \glsentryshort{dp-sgd} and evaluate the privatised versions (see Section~\ref{sec_dpsgd}).
    \item We publicly share our source code to facilitate reproducibility and future research\footnote{
    \ifreview
        https://github.com/ANONYMIZED for review
    \else
        \url{https://github.com/jesse-merhi/CNN-TRAJGAN}
    \fi
    }.
\end{enumerate}

This paper is organised as follows:
Section~\ref{sec_background} provides background on trajectory datasets, \gls{dp}, and \glspl{gan}.
Evaluation metrics are detailed in Section~\ref{sec_metrics}.
Section~\ref{sec_related_work} reviews related work.
Our objectives and contributions are outlined in Section~\ref{sec_problem_statement}.
The design of the \gls{rtct} is described in Section~\ref{sec_rtct}.
Section~\ref{sec_dpsgd} discusses the integration of \glsentryshort{dp-sgd}.
Our approach is evaluated in Section~\ref{sec_eval}, while Section~\ref{sec_discussion} addresses the results and potential future research directions.
The paper concludes with Section~\ref{sec_conclusion}.

\section{Background}\label{sec_background}
This section lays out the background knowledge for the remainder of this paper. 
Section~\ref{sec_trajectory} defines a trajectory dataset.
Section~\ref{sec_dp} provides an overview of \gls{dp}, emphasising its application in deep learning via \glsentryshort{dp-sgd}.
In Section~\ref{sec_gan}, we explore \glsentrylong{gan}s.
Section~\ref{sec_dcgan_bg} provides further details on the \gls{dcgan} architecture.

\subsection{Trajectory Datasets}\label{sec_trajectory}

A trajectory dataset consists of a number of trajectories: $D_T = \{T_1,\dots, T_n\}$.
Each trajectory $T_i$ itself represents an ordered sequence of locations $T_i = (l_{i1}, \dots, l_{in})$, where each location consists of multiple attributes.
We assume the minimal information content of a location to be spatial coordinates such as latitude and longitude, i.e., $l_{ij} = (lat_{ij}, lon_{ij})$. 
However, locations might be enriched with additional information.
For instance, some datasets record altitude \cite{Geolife1}, while others contain temporal information~\cite{Geolife1, fs_nyc}, or semantic information such as \glspl{poi}, i.e., the type of location.
Within this work, we assume spatial details to be minimally present. 
Temporal and semantic information are optional and can extend all proposed approaches.

\subsection{Differential Privacy}\label{sec_dp}

\glsreset{dp}
\textit{\gls{dp}}~\cite{Dwork2013} is a rigorous semantic privacy notion.
In contrast to syntactic privacy notions like $k$-Anonymity, \gls{dp} offers protection against any adversary independent of their background knowledge.
\gls{dp} is founded on the principle of \textit{plausible deniability}, i.e., participation in a query should not (significantly) affect the outcome.
Accordingly, participation in a dataset does not harm one's privacy. 
Mathematically~\cite{Dwork2013}:
\begin{definition}[Differential Privacy]
	A mechanism $\mathcal{K}$ provides $(\varepsilon, \delta)$-differential privacy if for all \textit{neighbouring} datasets $D_1$ and $D_2$, and all $S \subseteq Range(\mathcal{K})$ holds
	\begin{equation}
		\mathds{P}[\mathcal{K}(D_1)\in S] \leq e^{\varepsilon}\times \mathds{P}[\mathcal{K}(D_2)\in S] + \delta
	\end{equation}
\end{definition}
In the original definition, \textit{neighbouring} means that dataset $D_2$ can be obtained from $D_1$ by removing all user records $u$ from the dataset: $D_2 = D_1 \backslash  \{r_{xi} | x = u\}$.
In practice, different neighbourhood definitions, also called \textit{unit of privacy}~\cite{pets_sok}, are deployed. 
Most commonly, especially in machine learning, \textit{instance-level \gls{dp}} is used, which assumes that each user contributes exactly one record.
The primary privacy parameter, $\varepsilon$, typically ranges from $0.01$ to $10$ in general contexts~\cite{Erlingsson2014}, but may be higher in \gls{dl}~\cite{dpfyML}. 
The parameter $\delta$, representing the accepted failure probability, is often set at $\delta = \frac{1}{n}$ for $n$ records to ensure each record's protection, although some scenarios require lower values~\cite{McSherry2022}.
To meet these privacy requirements, \gls{dp} integrates noise addition through mechanisms such as Laplace, Gaussian, or exponential~\cite{Dwork2013, McSherry2008}.

\subheading{Differential Private Machine Learning}\label{sec_dp-sgd_bg}
Deep learning models often rely on sensitive data for training, raising privacy concerns. 
Consider a \gls{dl} model that targets the detection of diseases based on real medical data.
It must be impossible to derive training data from the model's outputs or from the model's parameters in case the trained model is shared.
However, different attacks~\cite{Shokri2017} demonstrate the privacy risk of deep learning models. 
To counteract this, differential privacy has been integrated into \gls{dl} model training~\cite{Abadi2016, dpfyML}.

\noindent\textbf{\gls{dp-sgd}}~\cite{Abadi2016} is the prevalent method for achieving \gls{dp} in deep learning~\cite{dpfyML}. 
To prevent privacy leakage, the influence of training samples on the model's parameters is bounded.
The first step in \gls{dp-sgd} involves clipping the gradients to a norm $C$, which limits the impact of each training sample.
Next, Gaussian noise is added to these clipped gradients, determined by the norm $C$ and a noise multiplier $\sigma$.
Finally, privacy accounting monitors the privacy budget $\varepsilon$ during training to ensure the model meets $(\varepsilon, \delta)$-\gls{dp} upon completion.

However, \gls{dp-sgd} only safeguards training data's influence on model parameters. 
Data used during prediction or generation in generative models remains unprotected. 
This limitation underpins our approach to designing a generative model that relies solely on noise during generation, as suggested in~\cite{pets_sok}. 
While privacy libraries simplify \gls{dp-sgd} usage, guaranteeing privacy for models like \lstmtrajgan, which require real data during generation, remains a challenge (\refer Section~\ref{sec_intro}).

\subsection{Generative Adversarial Networks}\label{sec_gan}

\glsreset{gan}
\textit{\glspl{gan}}~\cite{Goodfellow2014} are a \gls{dl} algorithm used in unsupervised machine learning. 
These networks consist of two main components: a generator ($G$) and a discriminator ($D$). 
$G$ generates data from random noise, aiming to mimic real data samples. 
$D$ assesses whether a sample is real or produced by $G$. 
This interaction forms a competitive framework, described by the adversarial loss:

\begin{equation}
\min_G \max_D \mathbb{E}_{x\sim p_{\text{data}}}[\log D(x)] + \mathbb{E}_{z\sim p_z}[\log(1 - D(G(z)))]
\end{equation}

In this equation, $p_{\text{data}}$ is the real data distribution, and $p_z$ is the distribution of the noise input for $G$. 
The training process refines $G$ and $D$ until $G$ can produce data almost identical to real samples. 
One advantage of traditional \glspl{gan} architecture in regard to privacy is that the generator receives only random noise as an input during generation.
Real data samples in \glspl{gan} are necessary only during the training phase.
This characteristic facilitates seamless integration with \gls{dp-sgd}. 
In contrast, accessing the real data during generation can lead to privacy leakage~\cite{pets_sok}.
Building on this understanding of \glspl{gan}, we explore the well-known \gls{dcgan} architecture next.

\subsection{DCGAN}\label{sec_dcgan_bg}
Most current \gls{gan} models are loosely based on \gls{dcgan}~\cite{DCGAN}, owing to its effectiveness in generating valid synthetic data across various applications~\cite{Chintala2023}. 
\gls{dcgan} is a \gls{gan} utilising convolutional layers in both its generator and discriminator that was developed for image generation.
Its five key architectural features include:\begin{itemize}
    \item Strided and transposed convolutions for downsampling and upsampling, replacing pooling layers.
    \item Batch normalisation in both generator and discriminator.
    \item \glsentryshort{relu} activation in the generator, except for \glsentryshort{tanh} in the output layer.
    Note that we used sigmoid for the output layer due to our normalisation to $[0;1]$.
    \item Leaky\glsentryshort{relu} activation in the discriminator, except for sigmoid in the output layer.
    \item Elimination of \gls{fc} hidden layers.
\end{itemize}
These modifications lead to \gls{dcgan}'s training stability and broad applicability.

\section{Metrics}\label{sec_metrics}

This section outlines the metrics for our utility evaluation.
We first discuss two metrics for evaluating spatial distribution: \gls{hd} (Section~\ref{sec_hd}) and \gls{wd} (Section~\ref{sec_wd}).
We then describe \gls{ttd} (Section~\ref{sec_ttd}) for sequential quality and introduce the novel \gls{trr} (Section~\ref{sec_trr}) to assess temporal properties.

\subsection{Hausdorff Distance}\label{sec_hd}\glsreset{hd}

The \textit{\gls{hd}} is a distance metric that quantifies the similarity between two sets of points.
It calculates the maximum distance from a point in one set to the nearest point in the other set.
Formally, for two point sets $A, B$ and a distance function $d$ \cite{hausdorff-distance}:
\begin{equation}
    HD(A, B) = \max\left\{\,\max_{a \in A} \min_{b \in B} d(a, b),\, \max_{b \in B} \min_{a \in A} d(a, b)\,\right\}
\end{equation}
This metric is widely used for evaluating spatial utility in trajectory protection and generation~\cite{Hua2015, Li2017, Chen2020, Rao2020, Ma2021}. 
We generate trajectories until their combined point count matches the real test set's point count. 
Then, we compute the \gls{hd} between both sets.
However, the \gls{hd} is significantly affected by outliers~\cite{pets_sok}, such that we also use the \gls{wd}.

\subsection{Wasserstein Distance}\label{sec_wd}\glsreset{wd}

The \gls{hd} is prevalent in trajectory comparison but is highly susceptible to outliers, where a single misplaced point can lead to a drastically different \gls{hd}.
To address this, we compare the point distributions directly as proposed in~\cite{pets_sok}. 
A key aspect is assessing if the generated data mimics the real data's distribution, such as urban concentration versus rural sparsity.
Intuitively, the \gls{wd} treats distributions as piles of dirt, measuring how much 'dirt' must be shifted to transform one distribution into another. 
Hence, this metric is also known as \gls{emd}.
However, standard \gls{wd} is limited to one-dimensional data, while spatial trajectories involve (at least) two dimensions (latitude and longitude). 
Therefore, we opt for the \gls{swd}, suitable for two-dimensional data. 
Given \gls{swd}'s computational overhead, we randomly draw \num{10000} locations from both the real and the generated dataset and compute the \gls{swd} between those sets.
Since these two metrics only evaluate the point distribution, we use the subsequently discussed \gls{ttd} for assessing the sequential characteristics.

\subsection{Total Travelled Distance}\label{sec_ttd}\glsreset{ttd}

The \textit{\gls{ttd}} represents the length of the entire trajectory.
For a trajectory $t = ((x_1, y_1), (x_2, y_2), \ldots, (x_n, y_n))$, the \gls{ttd} is defined as:
\begin{equation}
\text{\gls{ttd}}(t) = \sum_{i=1}^{n-1} \sqrt{(x_{i+1} - x_i)^2 + (y_{i+1} - y_i)^2}
\end{equation}
The \gls{ttd} is calculated by computing the total travelled distance for each trajectory in both the real and generated datasets, creating two sets of travel distances. We then use the \gls{wd} to compare these sets by measuring the discrepancy between their underlying distributions, quantifying how similar the generated data is to the real data regarding travel distances.

\subsection{Time Reversal Ratio}~\label{sec_trr}\glsreset{trr}

We identified a lack of metrics assessing the quality of the temporal features in generated trajectories.
To address this gap, we developed a new metric, \textit{\gls{trr}}, which assesses how accurately a trajectory represents the flow of time. The goal of \gls{trr} is to ensure that generated trajectories display minimal instances where time appears to move backwards, mirroring the consistent forward progression of time in real-life trajectories. 
Formally, for a trajectory $t = ((x_1, y_1, t_1), (x_2, y_2, t_2), \ldots, (x_n, y_n, t_n))$ with timestamps $t_{i}$, \gls{trr} is defined as:
\begin{equation}
\text{\gls{trr}} = \frac{\sum_{i=1}^{n-1} I(t_{i} > t_{i+1})}{n-1}
\end{equation}
Here, $I$ is the indicator function, equalling $1$ if $t_{i} > t_{i+1}$ (indicating a regression) and $0$ otherwise.
While minor errors in location points are tolerable for trajectory utility, generated trajectories exhibiting backward time movement are readily identifiable as synthetic. 
Therefore, optimising against such temporal inconsistencies is essential to enhance realism.

\section{Related Work}\label{sec_related_work} %

Numerous privacy-preserving mechanisms for trajectory data exist. 
Jin et al.~\cite{Jin2021} survey various approaches, including those based on $k$-Anonymity, while Miranda-Pascual et al.~\cite{Miranda-Pascual2023} focus on \gls{dp} mechanisms.
Despite significant progress, a privacy-utility trade-off persists \cite{Rao2020, Qu2020, Ma2021}. 
$k$-Anonymity offers practical privacy but lacks robust formal guarantees. 
Conversely, \gls{dp} mechanisms cause significant utility loss. 
Generating synthetic trajectories via deep learning emerged as a potential alternative \cite{Liu2018a}.
Notable implementations include the aforementioned LSTM-TrajGAN~\cite{Rao2020}.
Numerous other \gls{dl}-based generative models have been proposed~\cite{dp-trajgan, Jiang2023, Kim2022, lgan-dp, Ozeki2023, Shin2023, Song2023, TSG} and discussed in a recent SoK~\cite{pets_sok}.
Yet, none of the existing deep learning-based generative models can provide strong formal privacy guarantees, such as \gls{dp}~\cite{pets_sok}.
More recent approaches not considered in this SoK, such as DiffTraj~\cite{DiffTraj23}, improve utility further but still do not offer formal privacy guarantees and rely on the inherent privacy of generation instead.
However, generated data can leak information such formal guarantees remain desirable~\cite{pets_sok, PATE-GAN, Xie2018}.
Moreover, Buchholz et al.~\cite{pets_sok} mention that \glsentryshort{conv1d} layers appear to be superior for capturing the spatial distribution of trajectories compared to \glsentryshort{rnn}-based models.
This motivates our investigation of \gls{cnn}-based generative models.

\glsreset{ntg}
\textbf{LSTM-TrajGAN}\label{sec_lstm_trajgan} 
\cite{Rao2020, Rao2020_code} represents the most cited deep learning-based generative trajectory model.
The model consists of a generator and discriminator with similar architectures.
Real trajectories are normalised and encoded, combining location, temporal, and semantic features.
The generator embeds and fuses these encodings with noise into latent representations. 
This representation is processed by an \gls{lstm} layer, generating synthetic trajectories with multiple features. 
Unlike the classic \gls{gan} architecture (Section~\ref{sec_gan}), where the generator only receives Gaussian noise as input, LSTM-TrajGAN uses real trajectories as input during generation.
As pointed out in~\cite{pets_sok}, this has privacy implications and might lead to the generated trajectories leaking information about the input trajectories. 
Moreover, this architecture makes it challenging to integrate \gls{dp} guarantees.
Therefore, we use a variant of LSTM-TrajGAN, named \textbf{\gls{ntg}}~\cite{pets_sok}, as our evaluation baseline. Unlike the original, \gls{ntg} omits trajectory inputs during generation. Its architecture mirrors LSTM-TrajGAN's, with the primary difference being that its generator uses a noise vector alone.
We made the choice to use \gls{ntg} for two reasons.
First, \gls{ntg} can be trained with \gls{dp} guarantees by using \gls{dp-sgd} (\refer Section~\ref{sec_dp-sgd_bg}, which allows us two compare a \gls{dp} version of our \gls{poc} to \gls{dp}-\gls{ntg}.
Second, the architecture is more similar to \gls{dcgan}, which also uses noise-only input, such that the results focus more on the actual research question of comparing \gls{rnn} with \gls{cnn}-based models.

\textbf{Sequence Generation Based on CNN.}
The preference for \glspl{rnn} in trajectory generation stems from their suitability for sequential data. 
Yet, \gls{cnn}-based models have demonstrated effectiveness in sequential tasks. 
For example, WaveGAN~\cite{Donahue2019}, uses \glsentryshort{conv1d} for audio generation, and PAC-GAN~\cite{pacgan}, employs a 2D \gls{cnn} for network traffic generation.

\noindent\textbf{WaveGAN}\label{wavegan},
adapts \gls{dcgan} (\refer Section~\ref{sec_dcgan_bg}) to 1D data by using Conv1D layers for the generation of synthetic audio waves~\cite{Donahue2019}. 
Despite the sequential nature of audio waves, this model produces high-quality audio suitable for multimedia applications. 
WaveGAN surpasses earlier models like SampleRNN~\cite{Mehri2016} and WaveNET~\cite{wavenet}, highlighting the effectiveness of \glsentryshort{conv1d} layers. 
The model exploits the periodicity of sound waves, which the sliding convolutional kernel captures.
While trajectories contain regular patterns, they do not exhibit the same periodicity, limiting the model's applicability~\cite{pets_sok}.

\noindent\textbf{PAC-GAN}~\cite{pacgan}
employs a 2D \gls{cnn}-based \gls{gan} to generate synthetic network traffic packets. 
This approach proposes a transformation algorithm that converts hex-encoded packets into 2D matrices compatible with \glspl{cnn}. 
This representation yields an up to \SI{99}{\%} success rate for individual traffic type generations while an initial encoding only reached up to \SI{30}{\%}.
This outcome highlights the critical role of data representation in model performance. 
These examples motivate the design of a transformation enabling the usage of \gls{cnn}-based models for trajectory generation, opening avenues for future research. 

\section{Problem Statement}\label{sec_problem_statement}

As noted in Section~\ref{sec_intro}, location trajectories are valuable for analyses but contain sensitive information. 
Related work indicates a limiting privacy-utility trade-off in traditional protection methods. 
Therefore, the generation of synthetic data represents a promising alternative.
However, while achieving high utility, current models struggle to provide rigid privacy guarantees~\cite{pets_sok}.
Moreover, while most current methods rely on \glspl{rnn}, \gls{cnn}-based architectures have shown success in other sequential domains, and initial experiments with convolutional layers for trajectory generation yield promising results~\cite{pets_sok}.
This leads to our research question: 
\textit{Can a \gls{cnn}-based \gls{gan} produce high-quality synthetic trajectories?}
To the best of our knowledge, this is the first attempt to utilise a fully \gls{cnn}-based architecture for trajectory generation. 
Our primary research question is subdivided into four sub-goals:

\glsreset{rtct}
\noindent\goal{Transformation}{goal-transform}\textbf{:}
    Develop a \gls{rtct} algorithm that transforms location trajectories into \gls{cnn}-compatible inputs.
    This transformation represents the main contribution of this research.
  
\noindent\goal{Integration}{goal-dcgan}\textbf{:}
    Integrate our transformation with a standard \gls{cnn}-based \gls{gan}, specifically \gls{dcgan} (\refer Section~\ref{sec_dcgan_bg}), to assess its performance in a \gls{poc}.
    We have chosen an established \gls{gan} model to highlight the flexibility and potential of our proposed transformation.
    The development of a specialised \gls{cnn}-based \gls{gan} tailored for trajectory synthesis remains an objective for future work.

\noindent\goal{Differential Privacy}{goal-dp}\textbf{:}
    Apply \gls{dp-sgd} to both our model and the considered baseline model \gls{ntg} to determine the impact of formal privacy guarantees on model performance.
    To the best of our knowledge, this is the first work using \gls{dp-sgd} for privacy-preserving trajectory generation, providing formal privacy guarantees.

\noindent\goal{Evaluation}{goal-eval}\textbf{:}
    Evaluate the potential of \gls{cnn}-based \glspl{gan} for trajectory generation by conducting an extensive evaluation on two real-world datasets, \gls{fs} and Geolife, with the four metrics outlined in Section~\ref{sec_metrics}.
    Additionally, we compare our model's performance with the \gls{rnn}-based model \gls{ntg}, described in Section~\ref{sec_lstm_trajgan}. 
    Moreover, we compare the performance of both models trained with \gls{dp-sgd} to measure the impact of formal privacy guarantees on utility.

\section{Reversible Transformation Design}\label{sec_rtct}
\glsreset{rtct}\glsreset{poc}

\begin{figure*}[!ht]
\centering
  \includegraphics[width=\linewidth]{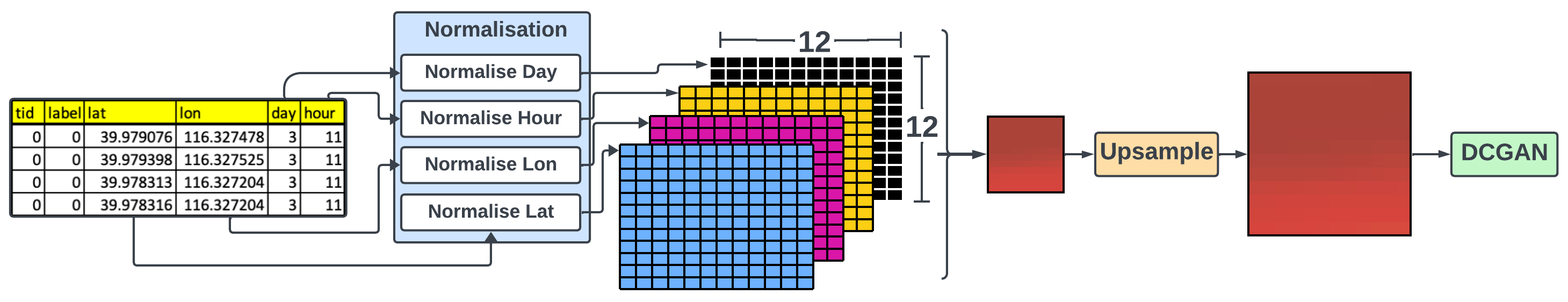}
  \vspace{-2em}
  \caption{
  \textbf{\gls{rtct}:} A trajectory's latitude, longitude, day, and hour values are normalised into a $12\times12$ matrix and upscaled to $24\times24\times4$ for \acrshort{dcgan}.
  }
  \label{fig_RCTC}
\end{figure*}

To enable the usage of \gls{cnn}-based \glspl{gan} for trajectory generation, we propose a \textit{\gls{rtct}} addressing \gref{goal-transform}.
This transformation, depicted in Figure~\ref{fig_RCTC}, comprises:
\begin{enumerate}
  \item Normalising the trajectories' latitude, longitude, day and hour values (Section~\ref{sec_normalisation}).
  \item Inserting these normalised values into a $12\times12\times3$ matrix (Section~\ref{sec_encoding}).
  \item Upsampling the resulting matrix to $24\times24\times3$ (Section~\ref{sec_encoding}).
\end{enumerate}
After generation, the synthetic trajectories are reverted into sequential form, as detailed in Figure~\ref{fig_reversion} and Section~\ref{sec_reversion}. 
Additionally, we demonstrate the potential of using \gls{cnn}-based models for trajectory generation by integrating our transformation with \gls{dcgan}~\cite{DCGAN} into a \gls{poc} implementation, addressing \gref{goal-dcgan} (Section~\ref{sec_integration}).

\subsection{Normalisation}\label{sec_normalisation}
Normalisation adjusts dataset features to a common scale, which is essential for deep learning. 
Typically, data is normalised to the range $[-1;1]$ or $[0;1]$ using techniques such as mean and standard deviation, and tanh normalisation~\cite{Singh2020}.
We employ min-max normalisation, which is reversible and scales data features into the range $[0; 1]$, making it ideal for the denormalisation of generated trajectories. 
For a feature $f$, with minimum value $min$ and maximum value $max$, the normalisation value $v$ is defined as:
\begin{equation}\label{eq_normalisation}
v = \frac{f - min}{max - min}    
\end{equation}
This normalisation method is applied to all features before using the trajectories for model training.

\subheading{Choice of max and min} 
Selecting appropriate maximum and minimum values for each feature is crucial for effective normalisation. 
Initially, we used global extremes for latitude and longitude: $-180^{\circ}$ to $180^{\circ}$ for longitude and $-90^{\circ}$ to $90^{\circ}$ for latitude.
This method proved too coarse, as slight changes in coordinates, which could represent significant distances, were not adequately captured. For example, a change from $(lat, lon) = (40, 70)$ to $(40.0001, 70)$ represents a distance of approx.\ $10$ meters but is negligible in global normalisation.
Moreover, the values have to be dataset-independent to prevent privacy leakage.
Finally, we opted for \textit{geographical constraints} tailored to the datasets, such as a ring road for a Beijing dataset or a bounding box aligned with a city's official boundaries.
This approach does not access the dataset but uses public knowledge, thus preventing information leakage.
For \textit{time-based features}, we applied fixed maxima and minima, $0\mbox{-}6$ for days and $0\mbox{-}23$ for hours, aligning with the natural constraints of these temporal features. This ensures sufficient variability for the model to detect patterns effectively.

\subsection{Trajectory Encoding}\label{sec_encoding}

Our trajectory encoding underwent two main iterations. 
Initially, we used an asymmetrical matrix as a "strip" to store trajectory information, but this format yielded unsatisfactory results. 
To overcome these limitations, we evolved our approach into a two-dimensional square convolution with multiple channels and clustered features. 
The final encoding scheme is depicted in Figure~\ref{fig_RCTC}.

\noindent\textbf{Strip Encoding}
Our initial encoding used a $144\times4$ strip-matrix, inspired by WaveGAN's audio encoding (see Section~\ref{wavegan}).
The matrix's first dimension, $144$, represents the maximum trajectory length, which depends on the dataset.
The second dimension captures features, here: latitude, longitude, day, and hour.
Early tests showed the model's learning was sub-optimal, possibly due to the lack of consistent periodicity in trajectories compared to audio waves.
Thus, we adopted a two-dimensional representation for each feature and a replication strategy
to reduce the effect of minor perturbations.

\begin{figure}[b]
\centering
  \includegraphics[width=\linewidth]{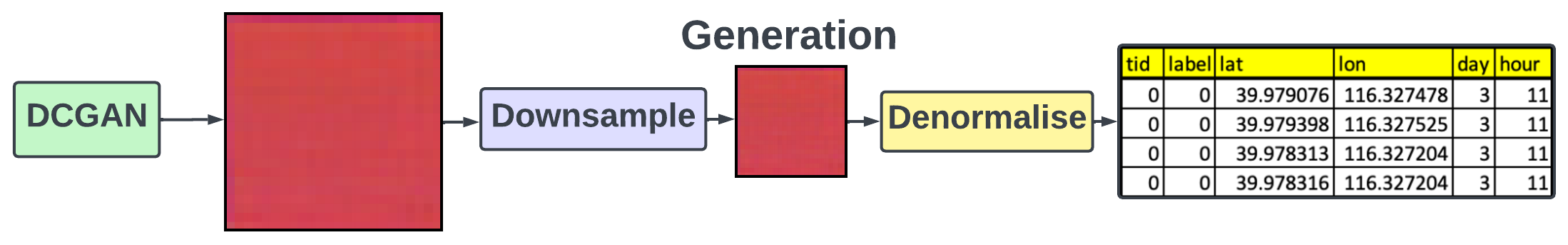}
  \caption{\textbf{Reversion:} \gls{dcgan} generates a $24\times24\times4$ trajectory, which is downsampled to $12\times12\times4$ and denormalised to retrieve the original format.}
  \label{fig_reversion}
\end{figure}

\noindent\textbf{Multi-Dimensional Square Convolution Encoding}
Our final encoding scheme evolved into a $12\times12\times4$ matrix, shown in Figure~\ref{fig_RCTC}.
It accommodates $144$ trajectory points, with each cell holding four separate feature channels.
This format generalises to trajectories with maximum length $x$ and $y$ features, resulting in a matrix size of $\lceil\sqrt{x}\rceil\times\lceil\sqrt{x}\rceil\times{y}$.
The main drawback of \gls{cnn}-based models over \gls{rnn}-based ones is the need to pre-define an upper trajectory length.
To mitigate perturbation effects, we upscaled our matrix to $24\times24\times4$, replicating features into $3$ additional cells~\cite{Goodfellow2014a}.

\subsection{Reversion}\label{sec_reversion}

The generator outputs synthetic trajectories in the introduced encoded format.
Therefore, the reversibility of this encoding into trajectories is essential.
Figure~\ref{fig_reversion} illustrates the reversion process.
Starting with a synthetic trajectory encoding of shape $24\times24\times4$, we downsample it to $12\times12\times4$. 
The downsampling divides the matrix into $2\times2$ squares, selecting one cell per square using the nearest-neighbour method.
After downsampling, values $v$ are denormalised into the feature $f$ using the rearranged normalisation Equation~\ref{eq_normalisation}: 
\begin{equation}
    f = v\times(max-min)+ min    
\end{equation}
This reversion is applied to each cell.
The features are then concatenated to revert to the original trajectory format.

\section{DCGAN Integration}\label{sec_integration}

As highlighted in Section~\ref{sec_dcgan_bg}, \gls{dcgan} is a popular \gls{gan} choice due to its ease of use and versatility. 
Therefore, we integrated \gls{dcgan} with \gls{rtct} as a \glsentrylong{poc}\footnote{
    The implementation is available at \ifreview
        https://github.com/ANONYMIZED for review
    \else
        \url{https://github.com/jesse-merhi/CNN-TRAJGAN}
    \fi
}. 
Although \gls{dcgan} was primarily designed for image generation and may not be ideal for trajectories, its adaptability makes it suitable for our \gls{poc} addressing \gref{goal-dcgan}, as this work focuses on transformation rather than the generative model itself.
In the following, we provide details on the implementation and optimisation strategies.

\subsection{Implementation}
We based our \gls{poc} on the \gls{dcgan} implementation from the PyTorch GAN repository~\cite{pytorch_gan}.
PyTorch data loaders facilitate dynamic access to data, crucial for training.
Our transformation is embedded in the data loader, managing loading raw trajectories from files to the encoding with \gls{rtct}.

\noindent\textbf{Padding and Masking.}
As described in Section~\ref{sec_encoding}, CNNs require constant length inputs, unlike RNNs.
Therefore, we pad trajectories shorter than the upper limit of $144$ points using $0$-post padding.
Masks based on the trajectories' original lengths are generated, with a vector for a trajectory of length $l$ comprising $l$ ones and $144-l$ zeros.
When applied to trajectories, this masking excludes zero-padding from affecting computations within the computational graph.

\subsection{Optimising DCGAN}\label{sec_optimisations}

To enhance the baseline PyTorch \gls{dcgan} performance, we employed several optimisation strategies:
\begin{enumerate}
    \item Implementing \gls{ttur}~\cite{TTUR}.
    \item Using a learning rate scheduler~\cite{learningrate}.
    \item Applying label smoothing~\cite{how_to_train_gan}.
\end{enumerate}

\glsreset{ttur}
The \textbf{\gls{ttur}}~\cite{TTUR} refers to using a lower learning rate for the generator than for the discriminator.
This optimisation facilitates the discriminator converging to a local minimum while the generator progresses more slowly~\cite{TTUR}.
We integrated \gls{ttur} through a \textit{generator factor} that reduces the generator's learning rate to one-tenth of the discriminator's learning rate.

\textbf{Learning rate schedulers} are widely used for optimising performance in deep learning~\cite{learningrate}.
While adaptive rates in optimisers like the Adam Optimiser~\cite{adam-optimizer} are common, we observed improved results by manually reducing the learning rates of both the discriminator and generator at specific milestones.
A learning rate scheduler speeds up initial learning and helps find a local minimum later, avoiding overstepping.

\textbf{Label smoothing}, our final optimisation, modifies trajectory labels to reduce overfitting~\cite{how_to_train_gan}. 
We adjusted valid labels to $0.9$ and fake labels to $0.1$.
This reduces the discriminator's confidence and results in better gradients during training, facilitating a smoother and more stable learning process~\cite{how_to_train_gan}.

\section{DP-SGD}\label{sec_dpsgd}
As discussed in Section~\ref{sec_related_work}, the core limitation of existing deep learning models for trajectory generation are the missing formal privacy guarantees. 
Today's de-facto standard is \gls{dp} described in Section~\ref{sec_dp}.
Despite efforts to deploy \gls{dp} for trajectory generation, studies \cite{Miranda-Pascual2023, pets_sok} show frequent misapplication, affecting the integrity of the privacy guarantees.
Thus, \cite{pets_sok} recommends using the established method of \gls{dp-sgd}, which ensures instance-level \gls{dp} if each training sample corresponds to one trajectory (see Section~\ref{sec_dp-sgd_bg}).

To address \gref{goal-dp}, we implemented \gls{dp-sgd} for our \gls{poc} and the baseline \gls{ntg} (\refer Section~\ref{sec_baseline}) to evaluate the impact of \gls{dp} and allow for comparative analysis.
We employed the Opacus~\cite{Opacus} library to integrate \gls{dp-sgd} due to its straightforward interface.
This established framework helps us avoid the common pitfalls of custom \gls{dp} implementations, which have compromised the integrity of privacy guarantees in other studies~\cite{pets_sok, Miranda-Pascual2023}.
Any layers incompatible with \gls{dp-sgd} were automatically replaced by Opacus' model fixer, in particular, \gls{dcgan}'s BatchNorm layers were replaced by GroupNorm.
Typically, \gls{dp-sgd} is applied to the discriminator in a \gls{gan}~\cite{Xie2018}.
This approach ensures that the generator also adheres to \gls{dp} through the post-processing property of \gls{dp} since it only receives indirect data access via feedback from the discriminator.
However, we aimed at enabling usage of the \gls{wgan-lp} loss, which is reported to yield better results than the standard adversarial loss~\cite{iWGAN, wgan_lp, pets_sok}. 
Opacus does not support \gls{dp-sgd} for models trained with gradient penalty at this time.
Moreover, the discriminator is updated more frequently than the generator when using \gls{wgan}~\cite{wgan} (usually $\approx5\times$ as often), such that adding noise to the discriminator would result in more noise being added by \gls{dp-sgd}.
Therefore, we decided to train the generator with \gls{dp-sgd} instead, i.e., the noise is added to the generator's gradients instead of the discriminator's gradients.
Utilising the \gls{mnist-seq} as a toy dataset~\cite{pets_sok}, we confirmed that the baseline model \gls{ntg} could produce samples of comparable quality with \gls{dp-sgd} applied to the generator as it did without \gls{dp-sgd}.
Due to a bug in the underlying framework, we had to use the standard \gls{wgan} loss instead of \gls{wgan-lp} for the \gls{dp} version of the baseline model \gls{ntg}.
The \gls{dp} version of our \gls{poc} was successfully trained with \gls{wgan-lp}.

To minimise the performance degradation caused by the noise from \gls{dp-sgd}, we adhered to the guidelines from Google's DP-fy ML paper~\cite{dpfyML}.
We set $\varepsilon = 10.0$, considered the upper limit for realistic privacy in deep learning~\cite{dpfyML}.
For a dataset with $n$ samples, we adopted $\delta = 1/ n^{1.1}$ as commonly recommended~\cite{McSherry2022, dpfyML}.
Compared to the non-\gls{dp} model, we increase both the batch size $b$ and the number of epochs $e$ by a factor $F = 10$ to keep the number of steps ($s = e \cdot \frac{n}{b}$) constant but increase the batch size which reduces the noise that is added~\cite{dpfyML}.
We selected a gradient clipping norm of $C = 0.1$.
Due to time constraints and the high computational cost of \gls{dp-sgd}, we could not complete a full ClipSearch~\cite{dpfyML} and learning rate sweep, but we verified that these heuristics yield good results on the \gls{mnist-seq} dataset.

\section{Evaluation}\label{sec_eval}
This chapter addresses goal \gref{goal-eval}.
Section~\ref{sec_preprocessing} details the datasets used and their preprocessing methods. Section~\ref{sec_baseline} introduces the baseline model, \gls{ntg}. Subsequent sections discuss evaluation results: Section~\ref{eval_param} covers parameter choices, Section~\ref{eval_setup} outlines the hardware setup, Section~\ref{eval_results} presents results from standard models, Section~\ref{eval_dp} focuses on outcomes from training with \gls{dp-sgd}, and Section~\ref{eval_quality} provides a qualitative analysis.

\begin{figure*}[!ht]
\centering
\iffinal
    \includegraphics[width=\linewidth]{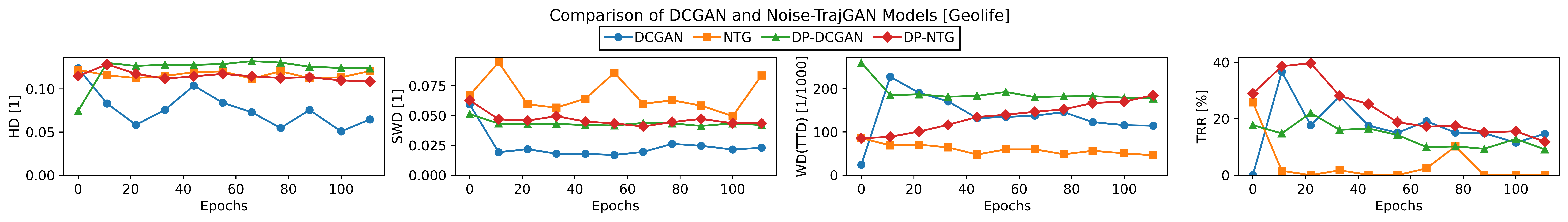}
\else
    \includegraphics[width=\linewidth]{img/metrics_geolife_small.png}
\fi
  \vspace{-2em}
  \caption{
  On Geolife~\cite{Geolife1}, \gls{dcgan} significantly outperforms \gls{ntg} for the spatial metrics, i.e., the model can capture the point distribution better.
  However, the \gls{rnn}-based \gls{ntg} excels at capturing the distance between consecutive points (\gls{ttd}). 
  \gls{dcgan} struggles to generate sensible timestamps (\gls{trr}).
  }
  \label{fig_results_geolife}
\end{figure*}

\subsection{Datasets}\label{sec_preprocessing}
To demonstrate the generalisability of our approach, we evaluate our implementation on two different datasets. 
First, the \gls{fs} dataset~\cite{fs_nyc}, used as a benchmark in \lstmtrajgan~\cite{Rao2020}, comprises $3,079$ trajectories with a maximum of $144$ locations each, primarily within New York City's bounds\footnote{
Bounding Box = $(40.6811, -74.0785)$ to $(40.8411, -73.8585)$
}.
We use the dataset as provided in \cite{Rao2020_code} without further preprocessing.
Second, the Geolife dataset~\cite{Geolife1} covers a larger geographical area around Beijing.
Moreover, \SI{91}{\%} of the dataset's trajectory have a sampling rate of $1$--\SI{5}{\second} or $5$--\SI{10}{\meter}~\cite{Geolife1} yielding more fine granular trajectories.
For consistency with \gls{fs}, we constrained Geolife's data within the fourth ring road of Beijing\footnote{
Bounding Box = $(39.8279,\ 116.2676)$ to $(39.9877,\ 116.4857)$
}.
We capped trajectories at $144$ locations and discarded those with fewer than $96$ locations to prevent extensive padding. 
This preprocessing reduced the available trajectories from \num{17621} to \num{7270}.

\subsection{Baseline Model}\label{sec_baseline}
LSTM-TrajGAN, as highlighted in Section~\ref{sec_lstm_trajgan}, is a leading generative model for trajectories.
Its use of real trajectory inputs during generation, however, raises privacy concerns and complicates the provision of \gls{dp} guarantees~\cite{pets_sok}.
To address these issues, we selected the noise-only variant \gls{ntg}, introduced in~\cite{pets_sok}, for its compatibility with \gls{dp-sgd} and its architectural similarity to our \gls{poc}, ensuring a fair comparison.
Despite \gls{ntg}'s inferior performance compared to LSTM-TrajGAN, it has the advantage that it integrates well with \gls{dp-sgd}, an essential feature for our comparative analysis.
We extended the model by training with \gls{dp-sgd} to facilitate comparisons of the \gls{dp} versions with our \gls{poc}.

\noindent\textbf{Hyperparameters} were generally maintained as per \cite{pets_sok}.
We trained \gls{ntg} using the \gls{wgan-lp} loss, which has been reported to yield the best outcomes~\cite{pets_sok}.
For the \gls{dp} version of the model, we had to revert to the standard \gls{wgan} loss due to an error caused by the combination of Opacus with the gradient penalty.
Adjustments were made to the learning rate, set at \num{1e-4}, an increased batch size of $64$ for alignment with our \gls{poc}, and extending the number of epochs to ensure a minimum of \num{10000} training steps, matching the step count of our \gls{poc}.
The parameters for the \gls{dp} version, referred to hereafter as DP-NTG, follow these modifications except for the changes outlined in Section~\ref{sec_dpsgd}.

\subsection{Hyperparameters}\label{eval_param}

This section outlines the training parameters used for the evaluation\footnote{
The repository contains detailed configuration files for all measurements.
}.
Both our \gls{poc} and the baseline model were trained for over \num{10000} steps for consistency, with a batch size of $64$.
The learning rate for our \gls{poc} was set to \num{2e-4}.
While we implemented the \gls{ttur} as described in Section~\ref{sec_optimisations}, we empirically determined that a generator factor of $1.0$, updating the generator and discriminator at the same frequency, yielded the best results.
A learning rate scheduler, referenced in Section~\ref{sec_optimisations}, reduces the learning rate by a factor of $0.1$ after $4\,000$ steps, and label smoothing is applied to the discriminator.
Contrary to \gls{ntg}, which employs the \gls{wgan-lp} loss, \gls{dcgan} showed superior performance with the standard adversarial loss (\refer Section~\ref{sec_gan}).
However, when training with \gls{dp-sgd}, the \gls{wgan-lp} loss with $5$ discriminator iterations per generator iteration proved superior for \gls{dcgan}.
The generator in our \gls{poc} accepts noise input shaped $(batch\_size, 100)$.
Network dimensions otherwise align with the \gls{dcgan} specifications from the PyTorch \gls{gan} repository~\cite{pytorch_gan}.
All remaining hyperparameters follow the PyTorch defaults.
Hyperparameters for the baseline model are outlined in Section~\ref{sec_baseline}.
During \gls{dp-sgd} training, most parameters were unchanged, with modifications to batch size, epochs, and learning rates as specified in Section~\ref{sec_dpsgd}.

\subsection{Evaluation Setup}\label{eval_setup}
All measurements were performed on a server (2x Intel Xeon Silver \num{4310}, \SI{128}{\giga\byte} RAM) with Ubuntu 22.04.4 LTS.  %
The server contains \num{4} NVIDIA GeForce RTX 3080 (\SI{10}{\giga\byte} RAM each), but only one GPU was used per experiment.  %
All measurements were executed with \num{5}-fold cross-validation.

\subsection{Results}\label{eval_results}

\begin{figure*}[!ht]
\centering
\iffinal
  \includegraphics[width=\linewidth]{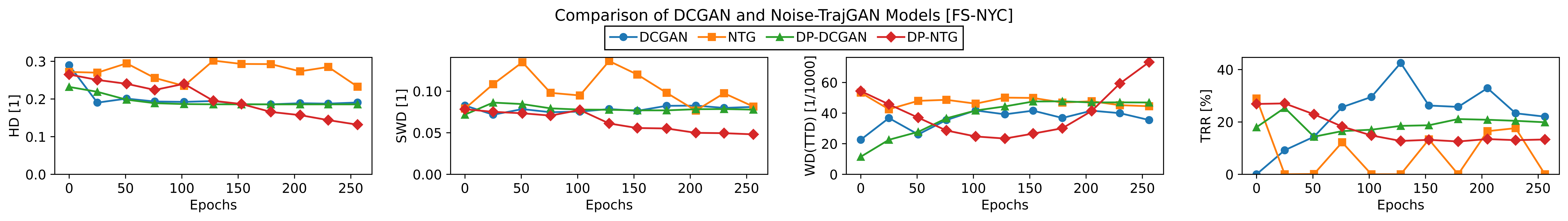}
\else
    \includegraphics[width=\linewidth]{img/metrics_fs_small.png}
\fi
  \vspace{-2em}
  \caption{
  On the \gls{fs}~\cite{fs_nyc} dataset, \gls{dcgan} captures the spatial distribution better than \gls{ntg}, although the difference is less significant than on Geolife.
  Interestingly, \gls{ntg} performs the worst in regard to the \gls{ttd} on \gls{fs}.
  The \gls{trr} shows similar results to Geolife.
  }
  \label{fig_results_fs}
\end{figure*}

This section presents a comparative utility evaluation between our \gls{poc} and \gls{ntg}.
The results of both models on the Geolife dataset are displayed in Figure~\ref{fig_results_geolife}, and those on \gls{fs} in Figure~\ref{fig_results_fs}.
Examples of one generated trajectory and the generated point distributions are provided in Figures~\ref{fig_progress} and Figure~\ref{fig_point_clouds}, respectively.
All results are reported at the end of the training, i.e., after approx. \num{100000} steps.

\subsubsection{Hausdorff Distance}
The \gls{hd} measures the maximum disparity between two sets of points (\refer Section~\ref{sec_hd}).
A lower \gls{hd} indicates closer similarity between the compared sets, thereby indicating better performance.
On Geolife, our approach achieved a \gls{hd} of $0.0646$.
\gls{ntg}, with a \gls{hd} of $0.120$, exhibited $1.87$ times worse performance.
On \gls{fs}, our \gls{poc} still performs better, but only by a factor of $1.45$. %

\subsubsection{Sliced Wasserstein Distance}
The \gls{swd} was employed as a second metric to evaluate the spatial distribution utility.
It assesses the shape, size, and density of the points within a generated dataset, with lower scores indicating a higher resemblance to the real dataset.
The results confirm the findings of the \gls{hd}, that our \gls{poc} is superior at capturing the spatial distribution of the dataset. 
On the Geolife dataset, \gls{dcgan} performs $3.64$ times better, and on \gls{fs}, $1.34$ times.

\subsubsection{Total Travelled Distance}
The \gls{ttd} (\refer Section~\ref{sec_ttd}) compares the distribution of travel distances in a generated dataset against the real-world dataset via the \gls{wd}.
Lower values indicate a greater resemblance to the actual dataset.
On the Geolife dataset, \gls{ntg} outperforms the \gls{poc} by a factor of $2.5$.
This outcome is expected, as the \gls{ttd} measures the distance between consecutive locations and depends on the model's ability to capture the sequential dependency of these locations.
Since \glspl{rnn} excel at capturing sequential properties, it appears reasonable that they outperform the \gls{cnn}-based model in this metric.
However, on the \gls{fs} dataset, \gls{ntg} performs significantly worse than \gls{dcgan} in terms of \gls{ttd}, which is surprising.
This variation could be attributed to the Geolife data consisting of relatively uniformly sampled trajectories, in contrast to the \gls{fs} dataset, which features user check-ins with greatly varied granularity and distances.

\glsreset{trr}\subsubsection{Time Reversal Ratio}
The \textit{\gls{trr}}, explained in Section~\ref{sec_trr}, counts the number of times two consecutive timestamps are impossible, i.e., an optimal score is $0$.
On both datasets, the \gls{poc} implementation performs worst with \SI{15}{\%} bad transitions on Geolife and \SI{22}{\%} on \gls{fs}, while \gls{ntg} produces barely any backward transitions.
As outlined for \gls{ttd}, the \gls{rnn} appears superior at capturing the dependency of a timestamp to its predecessor.

\subsection{DP Training}\label{eval_dp}

This section describes the evaluation of training with \gls{dp-sgd}.
The results for both models are displayed in Figures~\ref{fig_results_geolife} and~\ref{fig_results_fs}, and the generated point cloud of \gls{dp}-\gls{dcgan} is shown in Figure~\ref{fig_point_clouds}.
Initially, we conducted training on the toy dataset \gls{mnist-seq} and achieved results comparable to those of the model trained without \gls{dp-sgd}, using $\varepsilon = 10$ and the hyperparameter variations detailed in Section~\ref{sec_dpsgd}.
On Geolife, the \gls{dp} version of our \gls{poc} performs similar to \gls{ntg} in regard to the spatial metrics, but worse for the sequential metrics.
The point cloud visualisation (\refer Figure~\ref{fig_point_clouds}) shows that the produced points cloud looks nearly normally distributed, indicating a significant negative impact of the noise added to the gradients by \gls{dp-sgd}.
Yet, the centre of the distribution aligns with the highest density of real points, showing learning of the model.
On \gls{fs}, the results for the \gls{dp} version are similar to those of the standard \gls{poc}, which is a very promising outcome.
Nevertheless, \gls{dp-sgd} causes a significant utility degradation, and optimisation is required to make the \gls{dp} model practical.

Remarkably, DP-NTG appears to outperform its non-\gls{dp} counterpart in regard to the spatial metrics and even achieves the best values for both \gls{hd} and \gls{swd} on the \gls{fs} dataset.
Meanwhile, the sequential properties degrade substantially compared to the standard \gls{ntg} model.
On closer investigation, we noticed that the points generated by DP-NTG are randomly scattered over the entire map and do not exhibit any structure, which naturally yields very low \gls{hd} values. 
The outputs of DP-NTG appear to be mainly noise. 
We assume that the combination of the \gls{lstm} with \gls{dp-sgd} is highly unstable, as we encountered several issues during the evaluation and were unable to use \gls{wgan-lp}.
Moreover, the results show that DP-DCGAN and DCGAN perform similarly with DP-DCGAN being slightly worse, while the relationship between NTG and DP-NTG is less clear.
While our results are insufficient to make definitive claims, the results indicate that the combination of \gls{dp-sgd} with \gls{cnn}-based models might be preferable.
Our experiments show that applying \gls{dp-sgd} to these models is feasible but results in utility degradation.

\subsection{Qualitative Analysis}\label{eval_quality}

\begin{figure}[t]
\centering
\iffinal
  \includegraphics[width=\linewidth]{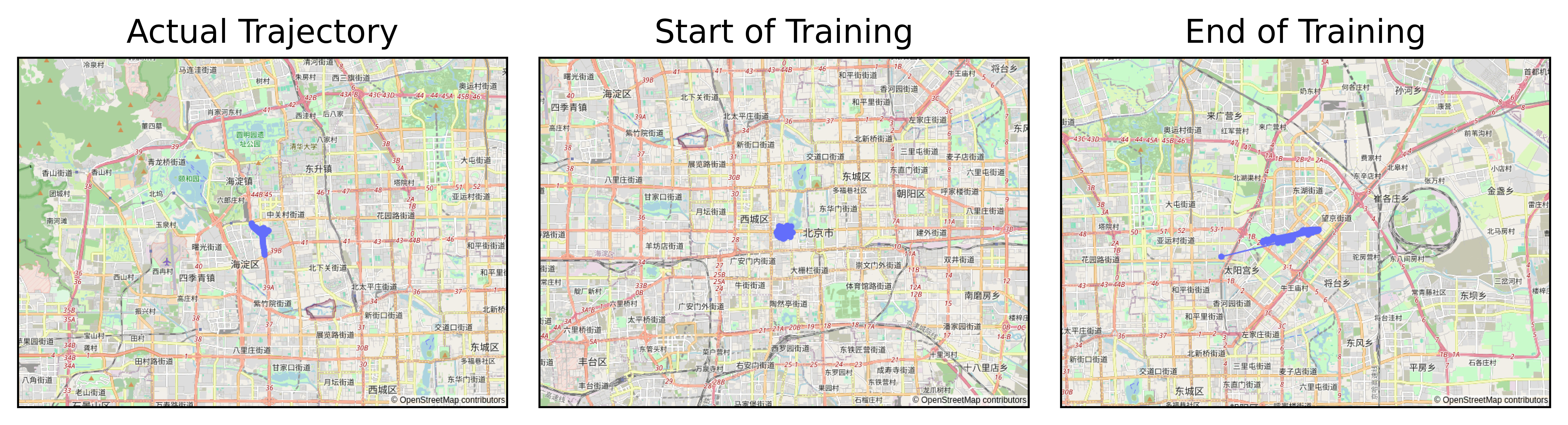}
\else
    \includegraphics[width=\linewidth]{img/progress_small.png}
\fi
  \caption{
  The real trajectory is dense and follows a primary direction.
  Generated trajectories are initially clustered in the map's centre but become more meaningful after training.
  }
  \label{fig_progress}
\end{figure}

Figure~\ref{fig_progress} displays an example of a generated trajectory, while Figure~\ref{fig_point_clouds} displays the distribution of generated locations over those in the real dataset.
While the generated data do not yet match the utility required for more complex downstream applications, they show a foundational ability to approximate spatial distributions. 
This aligns with the quantitative results described in the previous section.
The point clouds primarily capture the denser areas of the original dataset, demonstrating the model’s capability to learn essential spatial patterns.
However, finer details such as street layouts are not discernible, suggesting limitations in the current \gls{dcgan} architecture's capacity to model highly detailed urban geography.
A deeper network could improve on this shortcoming.

\section{Discussion}\label{sec_discussion}

This study's exploration into using \gls{cnn}-based \gls{gan}s for trajectory generation has yielded promising insights despite not achieving state-of-the-art results.
Our findings indicate that \gls{cnn}-based models are superior at capturing spatial distributions, whereas \gls{rnn}-based models better capture sequential properties, aligning with observations from related work~\cite{pets_sok}.
This difference in performance is likely attributed to the architecture of each model.
\gls{cnn}-based models find patterns in inputs using a sliding window that processes several inputs at once.
The \gls{rtct} transformation places values in a matrix, such that values within the same window are not necessarily sequentially adjacent.
Therefore, sequentiality is not necessarily the model's focus.
In contrast, \gls{rnn}-based layers always process values in a sequential manner, such that sequential dependencies are emphasised.
When comparing our prototype with state-of-the-art models such as LSTM-TrajGAN, it is important to note that these models exhibit higher utility because they modify real trajectories rather than generating new samples from Gaussian noise, as our \gls{poc} does.
While such approaches can improve utility, they may compromise privacy by handling real data directly. Our model, by generating synthetic trajectories, does not incur these privacy issues.
Additionally, the primary focus of our work was on the proposed \gls{rtct}transformation, hence the use of \gls{dcgan} "as-is" without tailoring it for trajectory generation.

While the prototype's ability to approximate spatial distributions is promising, its failure to capture finer details like specific street layouts underscores the limitations of the current \gls{dcgan} model.
This suggests a need for more sophisticated or larger network architectures to adequately capture the spatial patterns inherent in geographic data.
Similar challenges were noted by GeoPointGAN~\cite{GeoPointGAN}, which found that a standard PointNet could not capture minor patterns such as small roads, but using deeper networks improved results.
We plan to explore more complex \gls{cnn}-based models in future work, as it is beyond the scope of this paper.

Furthermore, we have demonstrated that training our \gls{poc} with \gls{dp-sgd} is feasible with acceptable overhead.
While the addition of noise impacts the results, this effect could be manageable for sufficiently large datasets and a privacy budget of $\varepsilon = 10$.
Considering the limited quality of the baseline results, more research is needed to fully assess the utility degradation caused by \gls{dp} training.
Nonetheless, \gls{dp-sgd} represents a promising direction for trajectory generation with formal privacy guarantees.
Overall, this project lays the groundwork for privacy-preserving, \gls{cnn}-based trajectory generation, which is discussed in the following section.

\subsection{Future Work}\label{sec_future}

The discussion highlights several avenues for future research.
The \gls{rtct} shows potential for adapting image-generation \glspl{gan} from computer vision to trajectory generation. This could allow the use of advanced generative models.

\begin{figure}[t]
\centering
\iffinal
    \includegraphics[width=\linewidth]{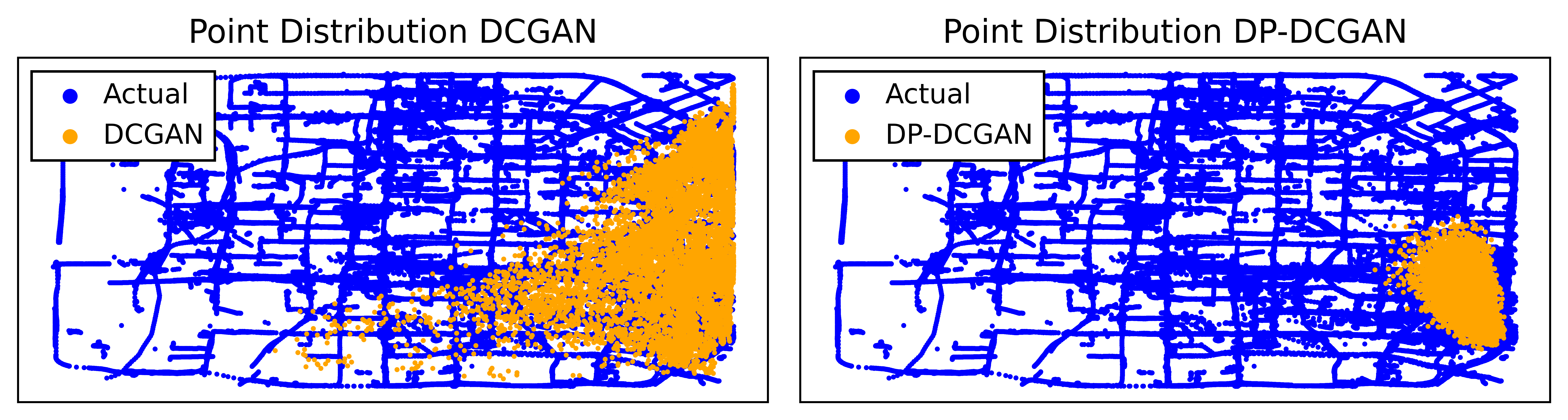}
\else
    \includegraphics[width=\linewidth]{img/point_clouds_small.png}
\fi
  \caption{
  The main density of points generated by \gls{dcgan} aligns with the dataset but lacks detail. The \gls{dp} version seems to be nearly normal distributed.
  }
  \label{fig_point_clouds}
\end{figure}

An important area for future development is a domain-specific \gls{cnn}-based \gls{gan}. 
Our current work has established a suitable encoding for \gls{cnn} models, but the next step is to refine the model architecture itself. 
From our discussion, several specific paths emerge.
First, exploring deeper, more complex models could enable capturing detailed aspects of datasets, such as the road network, beyond just the primary density clusters (\refer Figure~\ref{fig_point_clouds}).
A starting point for this direction could be the large PointNet proposed by GeoPointGAN for the generation of geographical point clouds~\cite{GeoPointGAN}.
Second, the model could focus more on the sequential properties, for instance, through combining a \gls{cnn} with an \gls{rnn} or positional encodings.
Third, future work could further examine the premise that \glspl{rnn} excel in sequential property representation while \glspl{cnn} are better suited for spatial distributions.
Fourth, implementing a specialised loss function akin to LSTM-TrajGAN's trajLoss could benefit model performance.

Additionally, future research could aim for a more equitable comparison between \gls{rnn}-based and \gls{cnn}-based \glspl{gan}. 
This could involve developing a \gls{cnn}-based \gls{gan} with an architecture similar to LSTM-TrajGAN, i.e., a \gls{cnn}-based model that receives a real trajectory encoding as input. 
Though a traditional \gls{gan} framework relying on noise generation is preferred for privacy, such an approach would facilitate a detailed assessment of the distinct strengths of \glspl{cnn} and \glspl{rnn} in trajectory generation.
Furthermore, concrete attacks and downstream applications could extend the privacy and the utility analysis, respectively.
For example, the \gls{tul}~\cite{marc2020} success rate could be determined as empirical privacy metric~\textbf{LSTM-TRAJGAN}, although an assignment of trajectory IDs to generated samples is not straightforward due to the noise-only architecture. 
In terms of downstream tasks, next week trajectory prediction~\cite{Fontana2023}, range queries~\cite{GeoPointGAN}, or hotspot preservation~\cite{GeoPointGAN} could be deployed.
Finally, future work could explore the impact of the privacy budget $\varepsilon$ on the provided utility and privacy by training multiple identical models and only varying $\varepsilon$. 

\section{Conclusion}\label{sec_conclusion}
Location trajectories, while valuable for various applications, inherently contain sensitive information, posing significant privacy concerns. 
Traditional trajectory privacy-protection mechanisms face a restrictive privacy-utility trade-off.
Therefore, deep learning-based generative models have been proposed as a promising alternative.
Yet, current models do not provide formal privacy guarantees, such as \gls{dp}.
Moreover, they mostly rely on \gls{rnn}-based architectures.
In this work, we 
\begin{enumerate*}
    \item introduced a \glsentrylong{rtct} enabling the usage of \gls{cnn}-based \glspl{gan} from computer vision for trajectory generation,
    \item integrated this transformation with the well-known \gls{dcgan} in a \glsentrylong{poc} implementation,
    \item integrated both this \gls{poc} and the baseline work \gls{ntg} with \gls{dp-sgd}, and
    \item evaluated the resulting four models across two datasets and four metrics.
\end{enumerate*}
To the best of our knowledge, this is the first instance of a fully \gls{cnn}-based \gls{gan} applied to trajectory generation and the initial application of \gls{dp-sgd} to ensure privacy in this context. 
Although our \gls{poc} reproduces major density patterns of the datasets, it fails to capture details and is less adept at capturing sequential and temporal patterns. 
While the \glsentrylong{poc} does not provide sufficient utility for real-world application, this development opens promising avenues for further research into the application of generative models from computer vision.
This study underlines the potential of \gls{cnn}-based generative models to either supplement or replace existing \gls{rnn}-based architectures in trajectory data privacy, encouraging further exploration in this promising research direction.

\section*{Acknowledgment}
\ifreview
    ANONYMISED
\else
    The authors would like to thank UNSW, the Commonwealth of Australia,
    and the Cybersecurity Cooperative Research
    Centre Limited for their support of this work.
    We thank the reviewers for their valuable feedback.
\fi

\bibliographystyle{IEEEtranModified}
\bibliography{library}

\ifprintglossary
    \appendix
    \printglossaries
\fi

\end{document}